\documentclass[twocolumn,superscriptaddress,pra,longbibliography]{revtex4-1}
\usepackage{graphicx}
\usepackage{amssymb}
\usepackage{amsmath}
\usepackage{epsfig}
\usepackage{color}
\usepackage{natbib}
\usepackage{mathtools}
\usepackage[colorlinks,linkcolor=blue,anchorcolor=blue,citecolor=blue,urlcolor=blue]{hyperref}

\begin{document}

\title{Optimal demonstration of Autler--Townes splitting}
\author{Yang Dong}
\author{Yu Zheng}
\author{Xiang-Dong Chen}
\author{Guang-Can Guo}
\author{Fang-Wen Sun}
\email{fwsun@ustc.edu.cn}
\affiliation{CAS Key Lab of Quantum Information, University of Science and Technology of China, Hefei,
230026, P.R. China}
\affiliation{Synergetic Innovation Center of Quantum Information $\&$ Quantum Physics, University of Science
and Technology of China, Hefei, 230026, P.R. China}
\date{\today }

\begin{abstract}
The atom-light interaction in a three-level system has shown significant physical phenomena, such as electromagnetically induced
transparency and Autler--Townes splitting (ATS), for broad applications in classical and quantum information techniques. Here, we optimally demonstrated the ATS with a quantum state manipulation method. The ATS in the dephasing-dominated diamond NV center system was successfully recovered by coherent microwave control, which cannot be observed with traditional method. The dynamical
process of ATS was investigated in detail, revealing a nontrivial quantum interference with geometric phase modulations. Based on the quantum interference, the signal of the optimal ATS is
twice as intense as those with traditional observation method. 
\end{abstract}

\maketitle
\section{Introduction}
Atom-light interaction is a fundamental topic in quantum optics and
atomic physics. In this realm, the optical response of the quantum multilevel
system can be dramatically modified  by quantum interferences among
different transition pathways, or by the strong Stark effect \cite{AT}. As a
typical representative of the former, electromagnetically induced
transparency (EIT) \cite{EIT1,EIT2,EIT3} can create an ultra-narrow
transparency window and delicately control the absorption and dispersion of
the medium, and thus many remarkable applications are being explored \cite%
{EIT1,EIT2,EIT3}.
The Stark effect also creates a transparency window because of the doublet
splitting structure in the absorption spectrum, which is called
Autler--Townes splitting (ATS). ATS has been employed to measure
transition dipole moments \cite{AAT1}, to control the spin-orbit
interaction in quantum system \cite{AAT2}, to suppress quantum decoherence \cite%
{AAT3,AAT4,AAT5,AAT7}, to dynamically control resonance fluorescence spectra
\cite{AAT6} and to create disorder for time crystals \cite{AAT8}. In the last
two decades, many systems have been used to investigate EIT
and ATS, such as atoms \cite{EITA1,AAT2}, super-conducting systems \cite%
{EITS1,ATS1,ATS2}, quantum dots \cite{EITQD2,ATQD1,CPTNV1,CPTNV2,CPTNV3}, defects in diamond \cite%
{EITNV1,AAT5,CPTNV1,CPTNV2,CPTNV3}, and nano-photonic systems \cite{EITO,EITP,EITM,EITM1,ATM1}.

Until now, almost all studies of EIT and ATS have been performed on
atomic-like system based on the traditional spectral-domain observational
method with long-duration driving (coupling and probe) pulses \cite%
{ATS1,ATS2,ATQD1,ATM1,AAT1,AAT2,AAT3,AAT4,AAT5,AAT6,AAT7}, wherein quantum
decoherence is dominated by the longitudinal relaxation process \cite%
{EIT1,EIT2,EIT3}. However, with recent developments in materials science
\cite{QDP1,EITS2,QDP2,QDP3}, rapid pure dephasing processes dominate
the quantum decoherence, such as in solid spin systems and superconducting
systems \cite{EITS2}. In these systems, when the driving pulses is
much longer than the dephasing time, the quantum coherence is lost and the
EIT and ATS phenomena disappear. Hence, the traditional observation method imposes a serious restriction on the investigation and application
of both phenomena. In this Letter, we optimally demonstrated the ATS by applying the quantum state manipulation \cite%
{QIP1,QIP2} method. The ATS was successfully recovered in diamond nitrogen vacancy (NV) center system where the quantum decoherence is
dominated by the dephasing process \cite{QDP1,EITS2,QDP2,QDP3}.

The diamond NV center has been one of promising
candidates for quantum information processes. Many studies have successfully demonstrated one-, and multi- qubit coherent operations \cite{NVONE1,NVTW1,NVTH1,NVM1} at room-temperature.  Using
the electron spin triplet state of the NV center, the $V$-type quantum
three-level system can be directly obtained. With the quantum state
manipulation method, ATS is observed in such a dephasing dominated system
and the dynamic process of ATS is investigated in detail by controlling the
pulse sequence. A nontrivial oscillation driven by the probe and coupling
field was experimentally revealed, including both geometric phase and
quantum interference in this three-level coupling. This dynamic behavior is notably different from that of a two-level system with Rabi oscillation.
Moreover, with delicately control of the interference and geometric
phase, the signal intensity of ATS is twice that the traditional method due to quantum interference, which is the optimal demonstration of ATS. 
\begin{figure}[tbp]
\centering
\includegraphics[width=8cm]{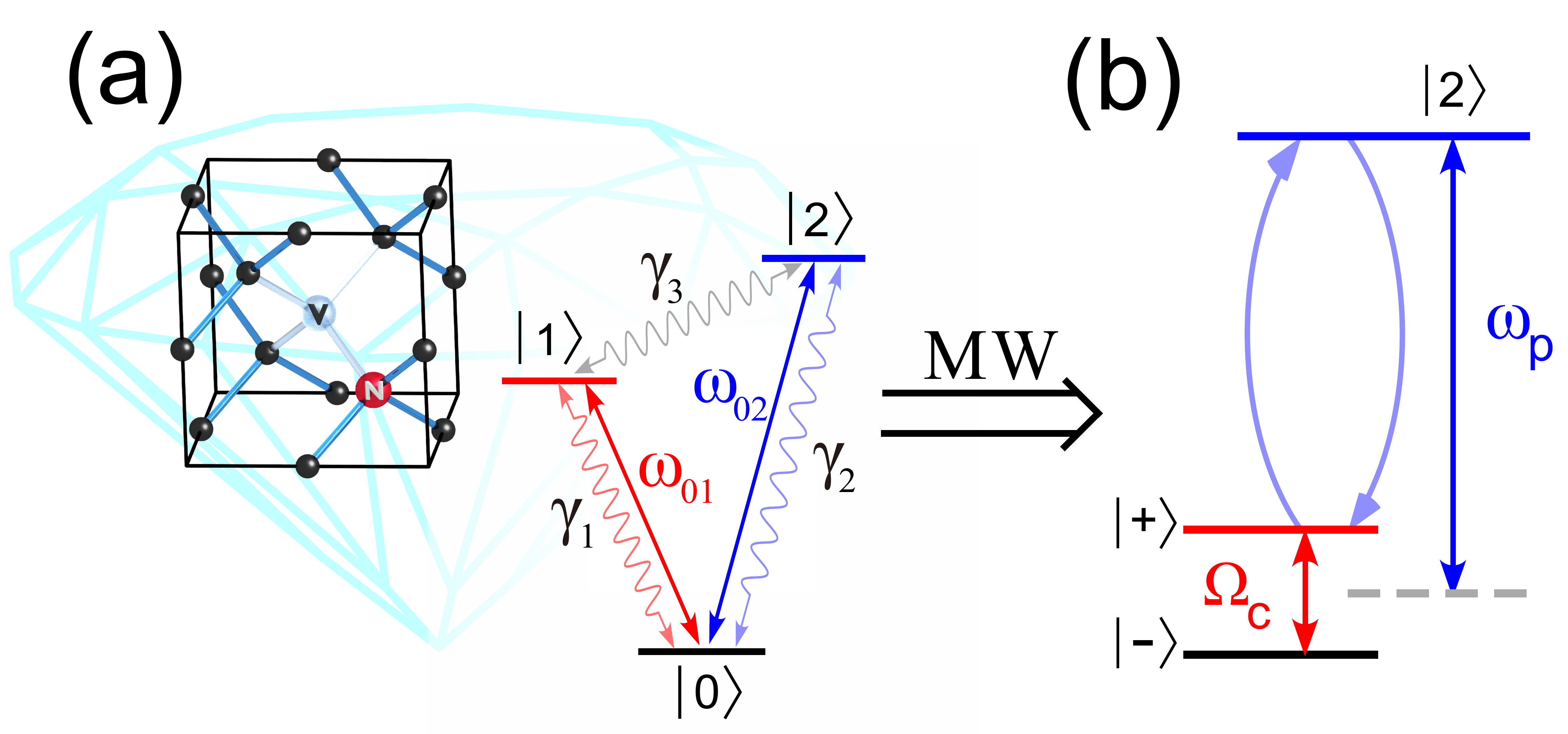}
\caption{(a) $V$-type three levels of the NV center ground state. The
degenerated ${m_{s}}=\pm 1$ sublevels are removed by applying a static magnetic
field. The dephasing rates of each level are $\protect\gamma _{1}$, $\protect%
\gamma _{2}$, and $\protect\gamma _{3}$. (b) Energy levels under a strong
coupling field. $\left\vert +\right\rangle =({\left\vert 0\right\rangle
+\left\vert 1\right\rangle )/}\protect\sqrt{2}$ and $\left\vert
-\right\rangle =$ $({\left\vert 0\right\rangle -\left\vert 1\right\rangle )/}%
\protect\sqrt{2}$ are separated by ${\Omega _{c}}$. The $%
\left\vert +\right\rangle \leftrightarrow \left\vert 2\right\rangle $ transition is
selected to study the dynamical process of the ATS.}
\label{fig1}
\end{figure}

\section{Recovery of ATS}
The NV center consists of a substitutional nitrogen atom adjacent to a
carbon vacancy, and the ground state exhibits zero-field splitting between
the ${m_{s}}=0$ and degenerate ${m_{s}}=\pm 1$ sub-levels of $D\approx 2.87$
$\mathrm{GHz}$ \cite{NVH,NVDNP}. The spin-dependent photon
luminescence (PL) enables the implementation of optically detected magnetic
resonance (ODMR) techniques
\cite{NVDNP} to detect the spin state with normalized $PL_{{m_{s}}%
=0}=1$ and $PL_{{m_{s}}=\pm 1}\approx 0.78$ in the current experiment. With secular approximation, the effective
Hamiltonian of the ground state triplet of the NV center \cite{NVH},
\begin{equation}
H=DS_{z}^{2}-{\gamma _{e}}B{_{z}}{S_{z}}\text{,}  \label{1}
\end{equation}%
is defined by the Zeeman splitting with the external magnetic field $B{_{z}}$
along the electron spin ${S_{z}}$, and ${\gamma _{e}}$ is the electron gyromagnetic
ratio. As a result, the $V$-type three-level system is formed with the states $%
\left\vert 0\right\rangle \equiv \left\vert {{m_{s}}=0}\right\rangle ,$ $%
\left\vert 1\right\rangle \equiv \left\vert {{m_{s}}=-1}\right\rangle $ and $%
\left\vert 2\right\rangle \equiv \left\vert {{m_{s}}=+1}\right\rangle $,
where ${\omega _{0,1}}$ (${\omega _{0,2}}$) is the transition frequency
between $\left\vert 0\right\rangle $ and $\left\vert 1\right\rangle $ ($%
\left\vert 2\right\rangle $), as shown in Fig.\ref{fig1}(a).

\begin{figure}[b]
\centering
\includegraphics[width=8cm]{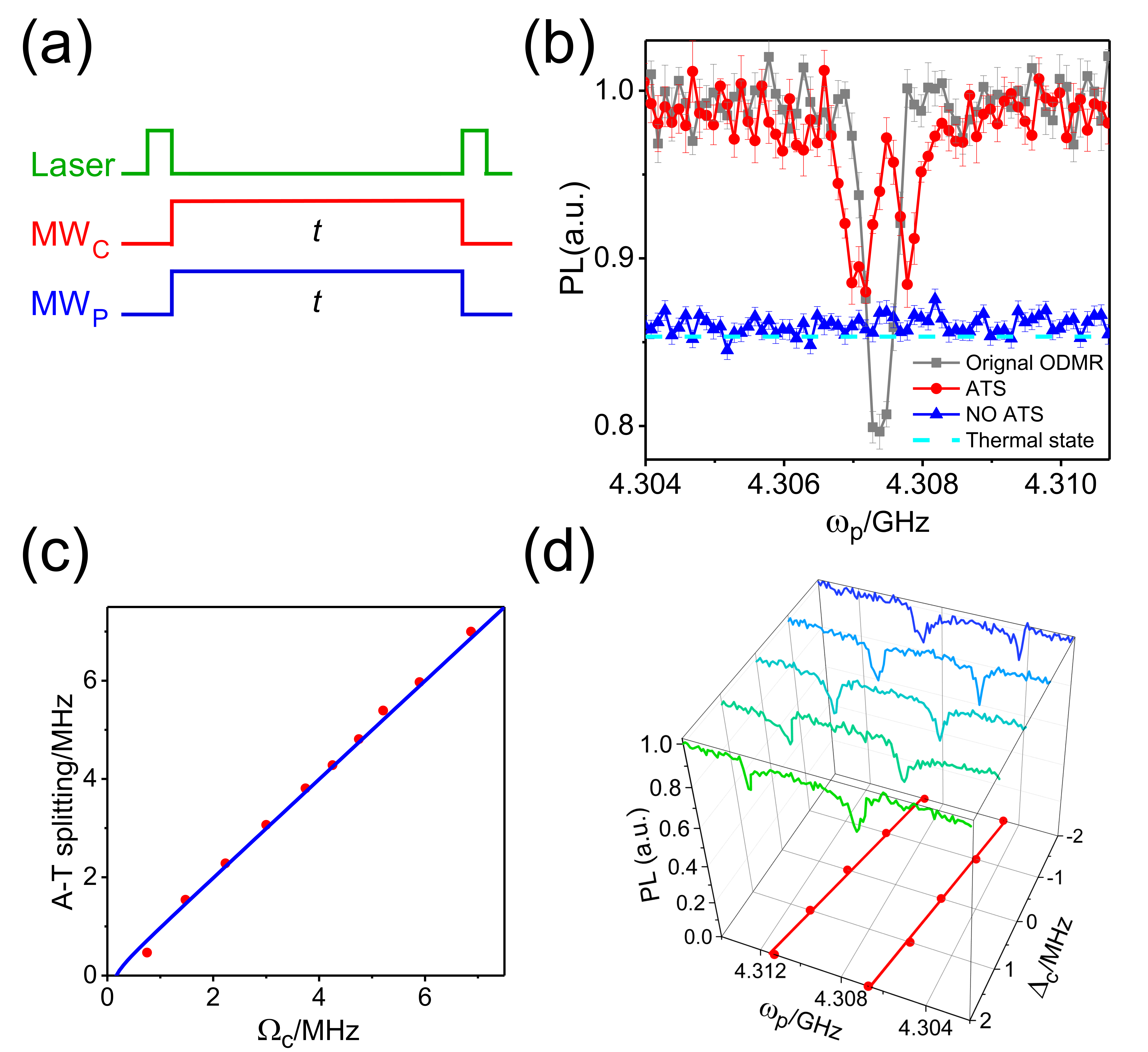}
\caption{(a) Optical and MW pulse sequences for the ATS experiment. Both
coupling and probe fields are simultaneously applied with identical durations.
(b) ODMR spectrum of the NV center. The blue triangles denote the
traditional measurement result with long time driving pulses. No splitting
was observed. The red dots denote the ATS. The cyan dashed
line shows the PL of the maximal mixed state of the NV center. The gray squares show the $\left\vert 0\right\rangle \leftrightarrow \left\vert
2\right\rangle $ transition, which demonstrates the original ODMR spectrum without a coupling field. (c) ATS versus
the amplitude of the coupling field with red dots. The blue curve is the
theoretical result. (d) ATS versus the detuning of the
couple field. In the bottom, the positions of the splitting are fitted by the
theoretical model, as shown with the red solid line. }
\label{fig2}
\end{figure}

When microwave (MW) coupling ($\omega _{c}$) and probe ($\omega _{p}$) fields are
applied to drive the NV center, the Hamiltonian of ATS with the
rotating-wave approximation is
\begin{equation}
{H_{ATS}}=\left[ {%
\begin{array}{ccc}
0 & \frac{{{\Omega _{c}}}}{2} & \frac{{{\Omega _{p}}}}{2} \\
\frac{{{\Omega _{c}}}}{2} & {\Delta _{c}} & 0 \\
\frac{{{\Omega _{p}}}}{2} & 0 & {\Delta _{p}}%
\end{array}%
}\right] \text{,}  \label{2}
\end{equation}%
where ${\Delta _{c}}={\omega _{c}}-{\omega _{0,1}}$ (${\Delta _{p}}={\omega
_{p}}-{\omega _{0,2}}$) is the frequency detuning between the coupling (probe) field
and the transition between $\left\vert 0\right\rangle $ and $%
\left\vert 1\right\rangle $ ($\left\vert 2\right\rangle $). Correspondingly,
${\Omega _{c}}$ and ${\Omega _{p}}$ are Rabi oscillation frequencies for the
coupling and probe fields, respectively.
When the coupling field is resonant with its corresponding transition (${%
\Delta _{c}=0}$) and much stronger than the probe field, the eigen-energy
levels are split by $\pm {\Omega _{c}}/2$ with eigenstates $\left\vert
+\right\rangle =({\left\vert 0\right\rangle +\left\vert 1\right\rangle )/}%
\sqrt{2}${\ and} $\left\vert -\right\rangle =({\left\vert 0\right\rangle
-\left\vert 1\right\rangle )/}\sqrt{2}$,
as shown in Fig.\ref{fig1}(b). For an atomic-like system, the quantum
decoherence is dominated by the longitudinal relaxation process such as
spontaneous radiation. The system will be in eigenstates with long-duration driving pulses. When the frequency of the probe field is
scanned, two absorption peaks can be observed after the probe field is
resonant with the eigenenergy levels \cite{AAT6,AAT7,Anisimov}, thus presenting ATS. This
is the traditional method based on the spectral measurement. However, for
single NV center in bulk diamond with the natural $^{13}C$ isotope, the dephasing process ($\sim 10$ $\mathrm{%
\mu s}$), which is caused by interaction with a nuclear spin bath \cite%
{NVL6,NVH}, is much faster than the longitudinal depolarization process ($%
\sim 2$ $\mathrm{ms}$) \cite{NVT12}. Thus when the driving
pulse is much longer than the dephasing time, the dephasing dominated system
would be in the maximally mixed state $\left( {\left\vert
0\right\rangle \left\langle 0\right\vert +\left\vert 1\right\rangle
\left\langle 1\right\vert +\left\vert 2\right\rangle \left\langle
2\right\vert }\right) /3$. The splitting would not be observed by scanning
the probe field (see the Appendix for the theoretical calculation and simulation.).

\begin{figure*}[t]
\centering
\includegraphics[width=17cm]{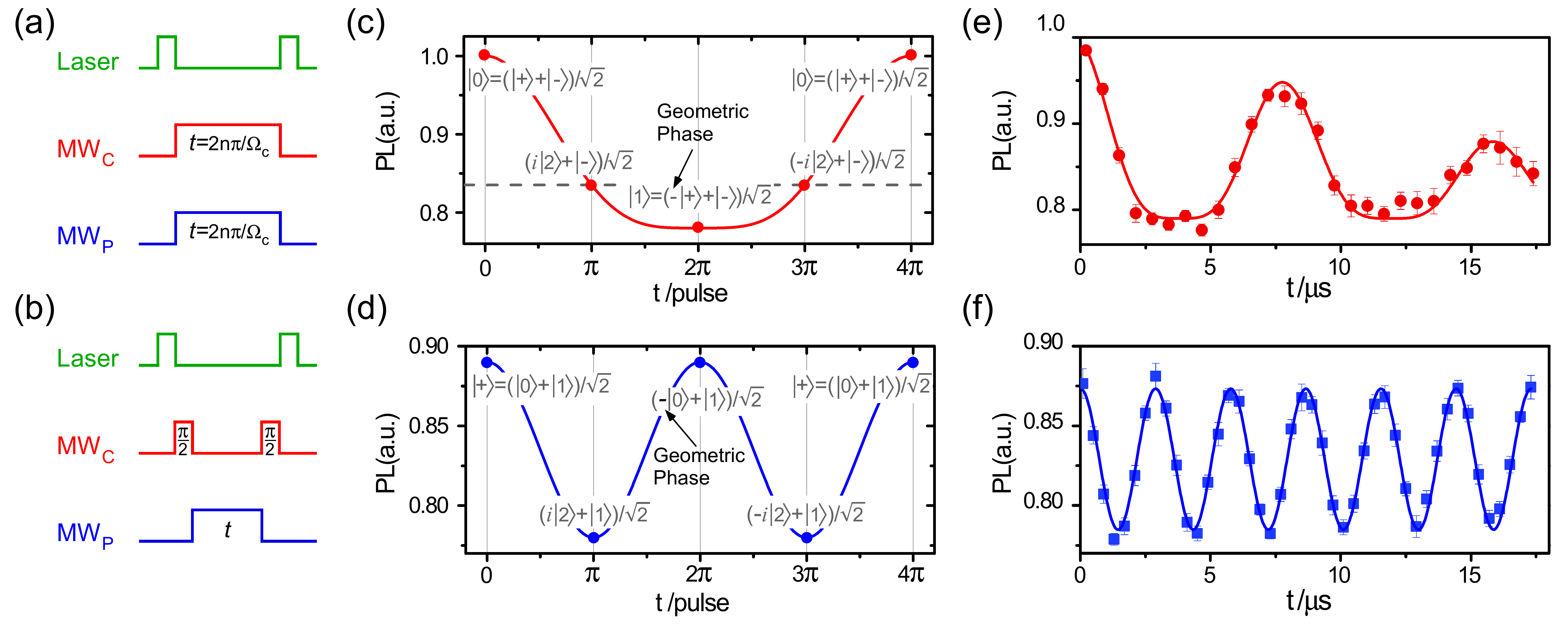}
\caption{(a) Optical and MW pulse sequences used to study the dynamical process
of ATS. Both driving fields were simultaneously applied, and the duration time was set to $t=2n\protect\pi /{\Omega _{c}}$, with ${\Omega _{c}}=2\pi  \times 4.73(1)$ $%
MHz $ and ${\Omega _{c}}/{\Omega _{p}}=14$. (b) Optical and MW pulse
sequences used for Rabi oscillation between ${\left\vert +\right\rangle
=(\left\vert 0\right\rangle +\left\vert 1\right\rangle )/}\protect\sqrt{2}$
and ${\left\vert 2\right\rangle }$. The $\protect\pi /2$ coupling pulse was
used to generate ${\left\vert +\right\rangle }$. The probe field was
resonantly applied to the transitions between ${\left\vert +\right\rangle }$ and ${%
\left\vert 2\right\rangle }$. (c)(d) The diagrams of the transition between ${%
\left\vert +\right\rangle }$ and ${\left\vert 2\right\rangle }$ with and
without ATS. (e)(f) Experimental results of the transition between ${%
\left\vert +\right\rangle }$ and ${\left\vert 2\right\rangle }$ with and
without ATS. The Rabi
frequency in (f) is $2.74(4)\approx 2\protect\sqrt{2}$ of that in (e).}
\label{fig3}
\end{figure*}

In this experiment, to fully study the ATS with NV center, we employ the
quantum state manipulation method, as shown in Fig.\ref{fig2}. The
electronic spin qubit was initialized into the ${m_{s}}=0$ state by a $3$ $%
\mathrm{\mu s}$ $532$ $\mathrm{nm}$ optical pulse. At the proper magnetic
field strengths ($51$ $\mathrm{mT}$), optical pumping also polarizes the $%
^{14}N$ nuclear spin of the NV center into ${m_{I}}=+1$, because resonant
polarization exchanges with the electron spin in the excited state. Then we simultaneously
applied coupling and probe fields with an identical duration time.
Finally, the electron spin state was read with another $532$ $\mathrm{nm}
$ optical pulse. When the duration time was set to be $t=52.2\mathrm{\mu s}$%
, which was much longer than the dephasing time, no splitting was observed
as the probe field was scanned. As shown in Fig.\ref{fig2}(b), the PL was
maintained at $\left( PL_{{\left\vert 0\right\rangle \left\langle 0\right\vert }}{%
+PL}_{{\left\vert 1\right\rangle \left\langle 1\right\vert }}{+PL}_{{%
\left\vert 2\right\rangle \left\langle 2\right\vert }}\right) /3\approx 0.85$%
, which corresponded to the maximally mixed state. This result demonstrates that quantum dephasing has a disastrous effect on the investigation of ATS
compare with the quantum longitudinal relaxation process.

However, if we control the duration time of the driving fields to $t=1.8\mathrm{%
\mu s}=\mathrm{\pi }/{\Omega _{p}}=2\mathrm{\pi }/{\Omega _{c}}$, the
doublet, which was spaced by the coupling Rabi frequency ${\Omega _{c}}$,
was observed to recover the ATS, as shown in Fig.\ref{fig2}(b). To explain this result, we present the ATS dynamics based on Eq.(\ref%
{2}) under the coupling field resonance with a single NV center, which is
expressed as \cite{AAT6,AAT7}

\begin{equation}
{H_{d}}=\left[ {%
\begin{array}{ccc}
\frac{{{\Omega _{c}}}}{2} & 0 & \frac{\sqrt{2}{{\Omega _{p}}}}{4} \\
0 & -\frac{{{\Omega _{c}}}}{2} & \frac{\sqrt{2}{{\Omega _{p}}}}{4} \\
\frac{\sqrt{2}{{\Omega _{p}}}}{4} & \frac{\sqrt{2}{{\Omega _{p}}}}{4} & {%
\Delta _{p}}%
\end{array}%
}\right] \text{,}  \label{3}
\end{equation}%
where $\left\vert +\right\rangle $, $\left\vert -\right\rangle $ and $%
\left\vert 2\right\rangle =\left\vert 2\right\rangle $ form the new basis
vectors. When ${\Delta _{p}}=0$, the effect of the probe field can be
neglected due to the large detuning between $\left\vert \pm \right\rangle $
and $\left\vert 2\right\rangle $. The spin state is in $\left\vert
0\right\rangle $ with the maximal PL. However, when ${\Delta _{p}}\approx
\pm {\Omega _{c}/2}$, we can eliminate the transition matrix element between
$\left\vert \mp \right\rangle $ and $\left\vert 2\right\rangle $ using the
second order perturbation theory. Hence, the probe field will drive the
system to oscillate between $\left\vert \pm \right\rangle $ and $\left\vert
2\right\rangle $ with lower PL, which demonstrates the ATS. Theoretically, the
splitting frequency from Eq.(\ref{3}) is ${\Delta _{AT}}={\Omega _{c}}%
+\Omega _{p}^{2}/(4{\Omega _{c}})$. In Fig.\ref{fig2}(c), we
independently measured the ATS as a function of the intensity of the
coupling field (denoted by Rabi frequency ${\Omega _{c}}$). Here, the
observed ATS is almost equal to the Rabi frequency of the coupling
field when ${\Omega _{c}\gg \Omega _{p}}$, which demonstrates the ATS characteristic. When the coupling field is not resonant, the doublet split dips are
not symmetric, which can be attributed to the unbalanced superposition of $%
\left\vert 0\right\rangle $ and $\left\vert 1\right\rangle $, and the
eigenenergies of the driven system are ${E_{\pm }}={\omega _{0,2}}+{\Delta _{c}%
}/2\pm {\Omega _{eff}}/2,$ where ${\Omega _{eff}}=\sqrt{\Delta
_{c}^{2}+\Omega _{c}^{2}}$. As shown in Fig.\ref{fig2}(d), when ${\Delta
_{c} \ll \Omega _{c}}$, the positions of ATS approximately have an linear relationship with the detuning of the coupling field when the effect of the Rabi frequency ($\Omega _{eff}$) remains unchanged.

\section{Dynamical process of ATS}
In addition to demonstrating splitting, we can study the dynamical process of ATS in detail. In the multi-level system, besides multiple separated transitions between different two levels, the quantum interference between those transitions shows a primary difference with the two-level system. Such quantum interference has been well demonstrated in EIT and usually believed to occur only in EIT. With the quantum state manipulation
method, the quantum interference was also revealed with geometric phase modulations in ATS.
In the experiment, the spin is first initialized into $\left\vert
0\right\rangle $, which is the superposition of $\left\vert +\right\rangle $
and $\left\vert -\right\rangle $, i.e. $\left\vert 0\right\rangle =({%
\left\vert +\right\rangle +\left\vert -\right\rangle )/}\sqrt{2}$. When the
probe field is resonant with the $\left\vert +\right\rangle
\leftrightarrow \left\vert 2\right\rangle $ transition by setting ${\Delta _{p}}\approx
{\Omega _{c}/2}$, it will flop the population between $\left\vert
+\right\rangle $ and $\left\vert 2\right\rangle $ at a frequency ${\Omega
_{+,2}}=\sqrt{2}{\Omega _{p}/2}$. When\ the state $\left\vert +\right\rangle $ is driven by
a $2\mathrm{\pi }$ probe pulse, it acquires a geometric phase $\mathrm{e}^{-%
\mathrm{i\pi }}$ \cite{NVTW1,NVG0,NVG1,NVG2}. Simultaneously, the state $\left\vert
-\right\rangle $ acquired a dynamical phase of $\mathrm{e}{^{\mathrm{i}{%
\Omega _{c}}t}}$ with the coupling field. If the pulse duration also
satisfies ${\Omega _{c}}t=$ $2n\mathrm{\pi }$ ($n=1,2,3,...$) for $\mathrm{e}%
{^{\mathrm{i}{\Omega _{c}}t}=1}$, the spin state is at $(-{\left\vert
+\right\rangle +\left\vert -\right\rangle )/}\sqrt{2}=-{\left\vert
1\right\rangle }$. In this case, another $2\mathrm{\pi }$ probe
pulse would be required to reconvert to $\left\vert 0\right\rangle $, as shown in Fig.%
\ref{fig3}(c) with the pulses sequences in Fig.\ref{fig3}(a).
Hence, the geometrical phase doubles the driving duration time
and halves the transition frequency. For comparison, Fig.\ref{fig3}(d)
shows the Rabi oscillation between $\left\vert +\right\rangle
\leftrightarrow \left\vert 2\right\rangle $ without the coupling field. In
experiment, when the electron spin state is detected by the operator $\left\vert
0\right\rangle \left\langle 0\right\vert $ with the ODMR method, Fig.\ref{fig3}%
(c) and (d) also show the PL of the two dynamic processes. The Rabi frequency oscillation is ${\Omega _{p}}$, whereas the ATS frequency is ${\Omega
_{+,2}/2=}\sqrt{2}{\Omega _{p}/4}$. This result can also be obtained by
solving Eq.(\ref{3}), where the population of $\left\vert 0\right\rangle
\langle 0|$ can be expressed as follows,
\begin{equation}
{P_{\left| 0 \right\rangle \langle 0|}} = \frac{1}{4}{\left| {\cos (\sqrt 2 {\Omega _p}t/4) + {{\text{e}}^{i{\Omega _c}t}}} \right|^2}.
\end{equation}%
This expression clearly demonstrates the quantum interference between $\left\vert
+\right\rangle \leftrightarrow \left\vert 2\right\rangle $ and $\left\vert
-\right\rangle $. In the experiment to study the quantum interference with
geometric phase modulations in ATS, we set $t=2n\mathrm{\pi }{/\Omega _{c}}$
with ${\Omega _{c}/\Omega _{p}=14}$ and detected the spin-dependent PL. The
experimental result is illustrated in Fig.\ref{fig3}(e). Because of the
dephasing processes, the data can be fitted by a theoretical curve $S(t)=a%
\mathrm{e}{^{-{{(t/T)}^{k}}}}\cos {^{4}}{[ w{(t-t_{c})}]}+b$, where $%
a=0.211(9)$, $T=17.5(1)$ $\mathrm{\mu s}$, $k=1.5(3)$, $t_{c}=7.85(5)\mathrm{%
\mu s}$, $b=0.790(3)$ and the frequency $w = 2\pi  \times 0.123(2){\text{MHz}}$. Correspondingly, the results of Rabi oscillation between $%
\left\vert +\right\rangle \leftrightarrow \left\vert 2\right\rangle $ is
shown in Fig.\ref{fig3}(f) with the Rabi frequency of $\Omega _{p}=2\pi  \times 0.338(1)$ $%
\mathrm{MHz}$. We can find $\Omega _{p}/\omega = 2.74(4)\approx 2\sqrt{2}$,
which is consistent with the above theory.

\begin{figure}[tbp]
\centering
\includegraphics[width=8cm]{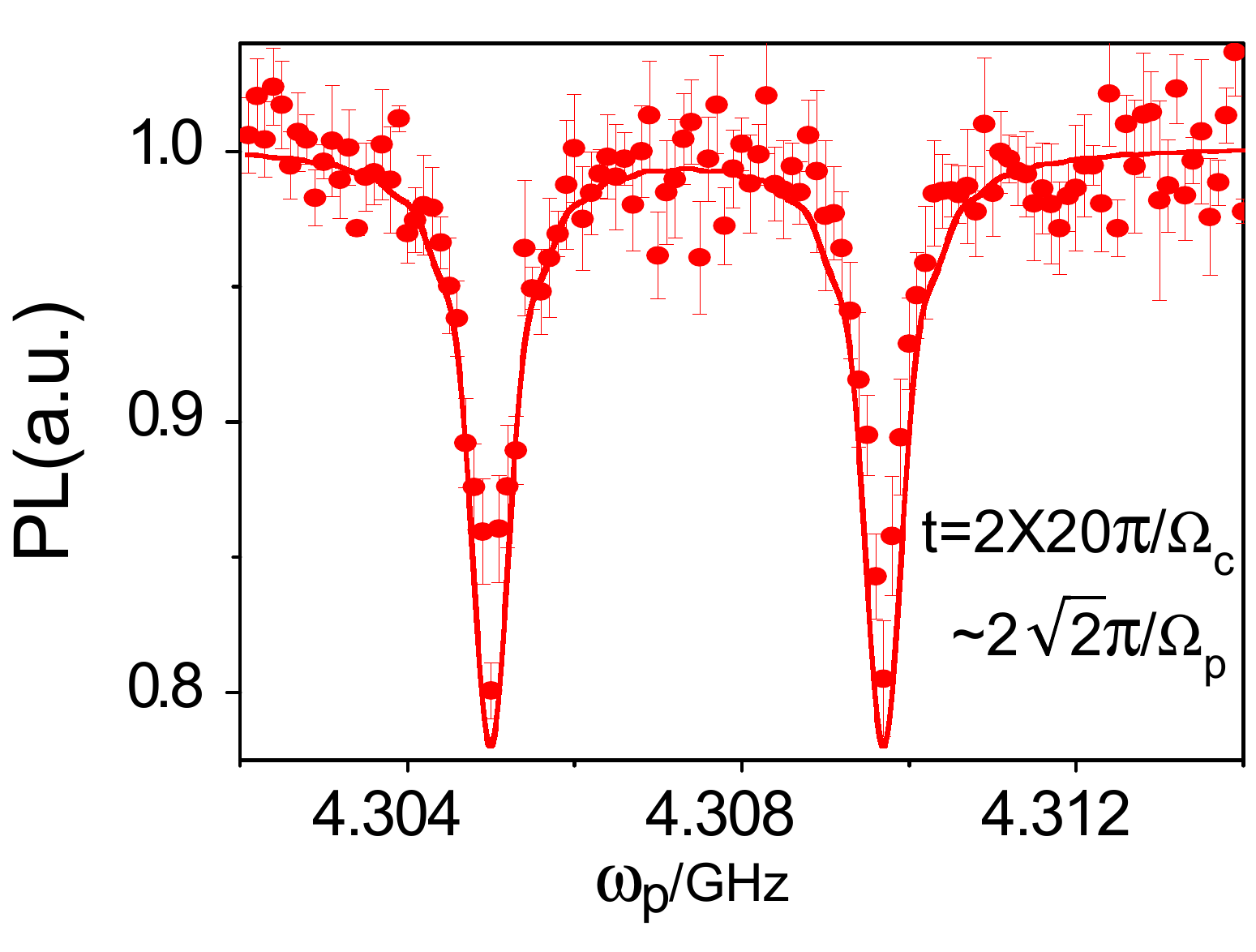}
\caption{Optimal demonstration of ATS with $t=2{%
\times 20}\protect\pi /{\Omega _{c}\approx }2\protect\sqrt{2}\protect\pi /{%
\Omega _{p}}$. The signal intensity (red dots) of the doublet dips is almost
same as the original signal in Fig.\protect\ref{fig2}(b).}
\label{fig4}
\end{figure}

\section{Optimal demonstration of ATS.}
The quantum state manipulation method is an optimal way to present the ATS with maximal signal intensity. Because of the quantum interference between $\left\vert
+\right\rangle \leftrightarrow \left\vert 2\right\rangle $ and $\left\vert
-\right\rangle $, the ATS signal can be optimized when the duration time
satisfies
\begin{equation}
{\Omega _{p}}t =2\sqrt{2}(2k-1)\mathrm{\pi }\text{, }
{\Omega _{c}}t =2{n}\mathrm{\pi }\text{,}
\end{equation}
or
\begin{equation}
{\Omega _{p}}t =2\sqrt{2}(2k)\mathrm{\pi }\text{, }
{\Omega _{c}}t =(2n-1)\mathrm{\pi }\text{,}
\end{equation}
with $n,k=1,2,3,\cdots$. In this case, the system remains in $\left\vert 1\right\rangle $, the
NV center provides minimal photon luminescence, and the ATS exhibits the
maximal dips. In the experiment, we set $t=2{\times 20}\pi /{\Omega _{c}\approx }%
2\sqrt{2}\pi /{\Omega _{p}}$ with ${\Omega _{c}/\Omega _{p}=14}$. By
scanning the probe field, we can obtain the result in Fig.\ref{fig4}. The
depth of the ATS is almost same as original signal in Fig.\ref{fig2}(b),
because of the quantum interference of the three-energy-level systems. In contrast, for the
traditional method \cite%
{ATS1,ATS2,ATQD1,ATM1,AAT1,AAT2,AAT3,AAT4,AAT5,AAT6,AAT7}, the signal
intensity of ATS is only half of the original signal for lacking quantum
interference with the third energy level. Therefore, such an enhancement in the signal with full control of quantum state in ATS may contribute to the precision measurement of the spectrum of a quantum system.

\section{Discussion}
In conclusion, we have presented an optimal observation method based on quantum state manipulation to study and demonstrate ATS. The ATS was recovered in a dephasing-dominated quantum
system, which can not be observed with traditional observation methods. With
the quantum state manipulation methods, the dynamical process of ATS was
investigated in detail with a nontrivial behavior from the quantum
interference with geometric phase modulations.
Consequently, the ATS was
optimally demonstrated, and its signal intensity was twice those of
other systems observed with the traditional observation method. The study presents a
feasible method to optimally observe the atom-light interaction in a multi-level system, which can be applied to investigate
quantum optics and atomic physics for a broad applications in high-dimensional quantum control and quantum error correction beyond the dynamically decoupling, decoherence-free subspace.
\section*{ACKNOWLEDGMENT}
This work was supported by the National Key Research and
Development Program of China (No. 2017YFA0304504), and the National Natural
Science Foundation of China (Nos. 11374290, 61522508, 91536219, and
11504363).

\appendix
\section{Experimental setup and work point of the quantum system}
\subsection{Experimental setup}
\begin{figure}[b]
\centering
\includegraphics[width=7.5cm]{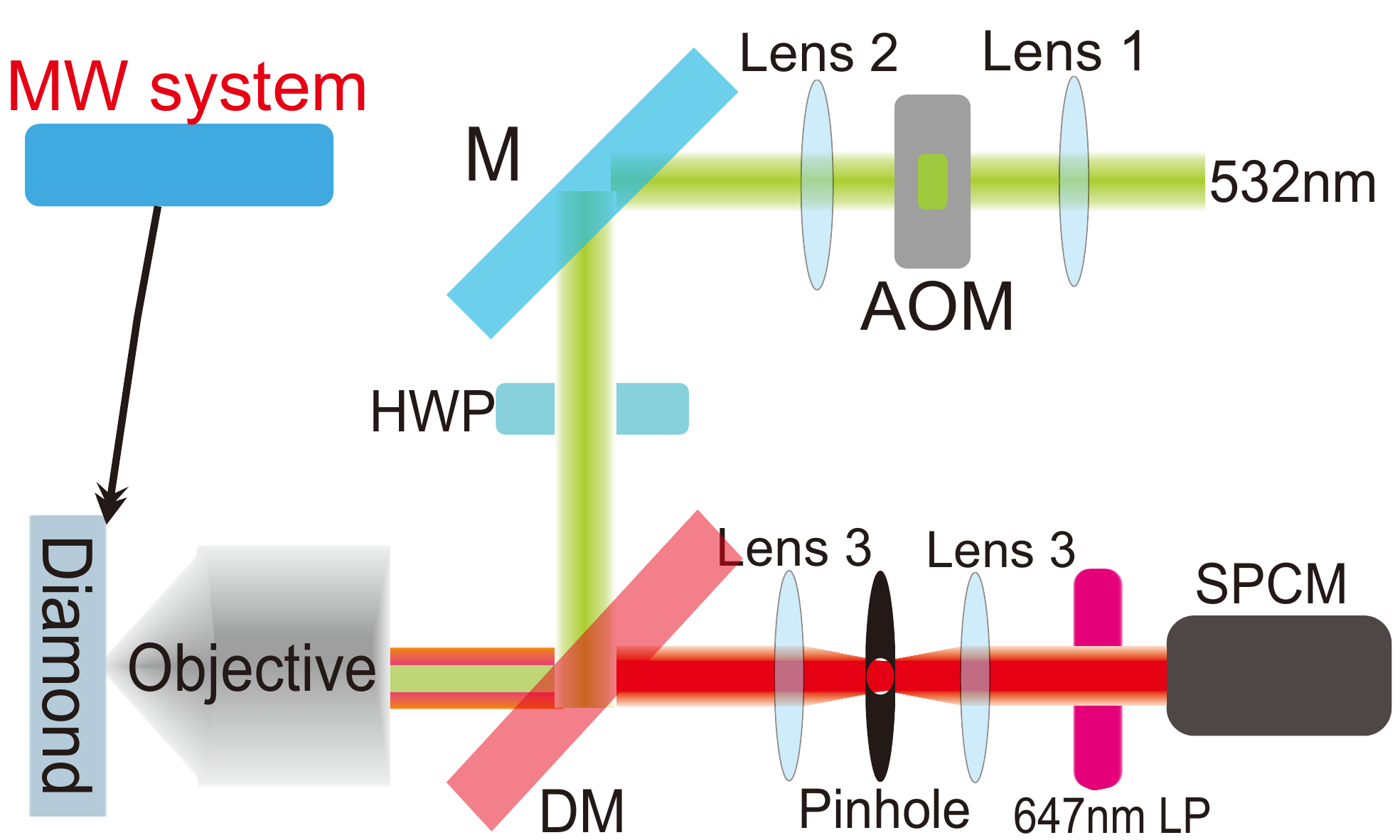}
\caption{Sketch of the experimental setup. Lens: optical lens with $f_1=100$ $mm$, $f_2=300$ $mm$ and $f_3=30$ $mm$; AOM:  acoustic optical modulator; M: mirror; HWP: half-wave plate for $532 $ $nm$ laser; DM: long pass dichroic mirrors (DM) edge wavelength $536.8$ $ nm$. The pinhole ($d=15 $ $\mu m$) and $647 $ $nm$ long pass filter were used to filter stray light; SPCM: single photon counting module.}
\label{SFig1S}
\end{figure}

As shown in Fig.\ref{SFig1S}, the NV center was located and detected with a home-built
confocal microscopy with a dry objective lens ($N.A. = 0.95$) at room temperature. The power of $532$ $nm$ continuous laser was set at $0.6$ $ mW$. The NV center fluorescence was separated from the excitation laser with a $647$ $nm$ long pass filter and then detected by single photon counting modules. We constructed two synchronized microwaves that drove the NV center system with two different frequencies. The microwave was coupled to the sample by a coplanar waveguide.\par
\begin{figure}[tbp]
\centering
\includegraphics[width=7.5cm]{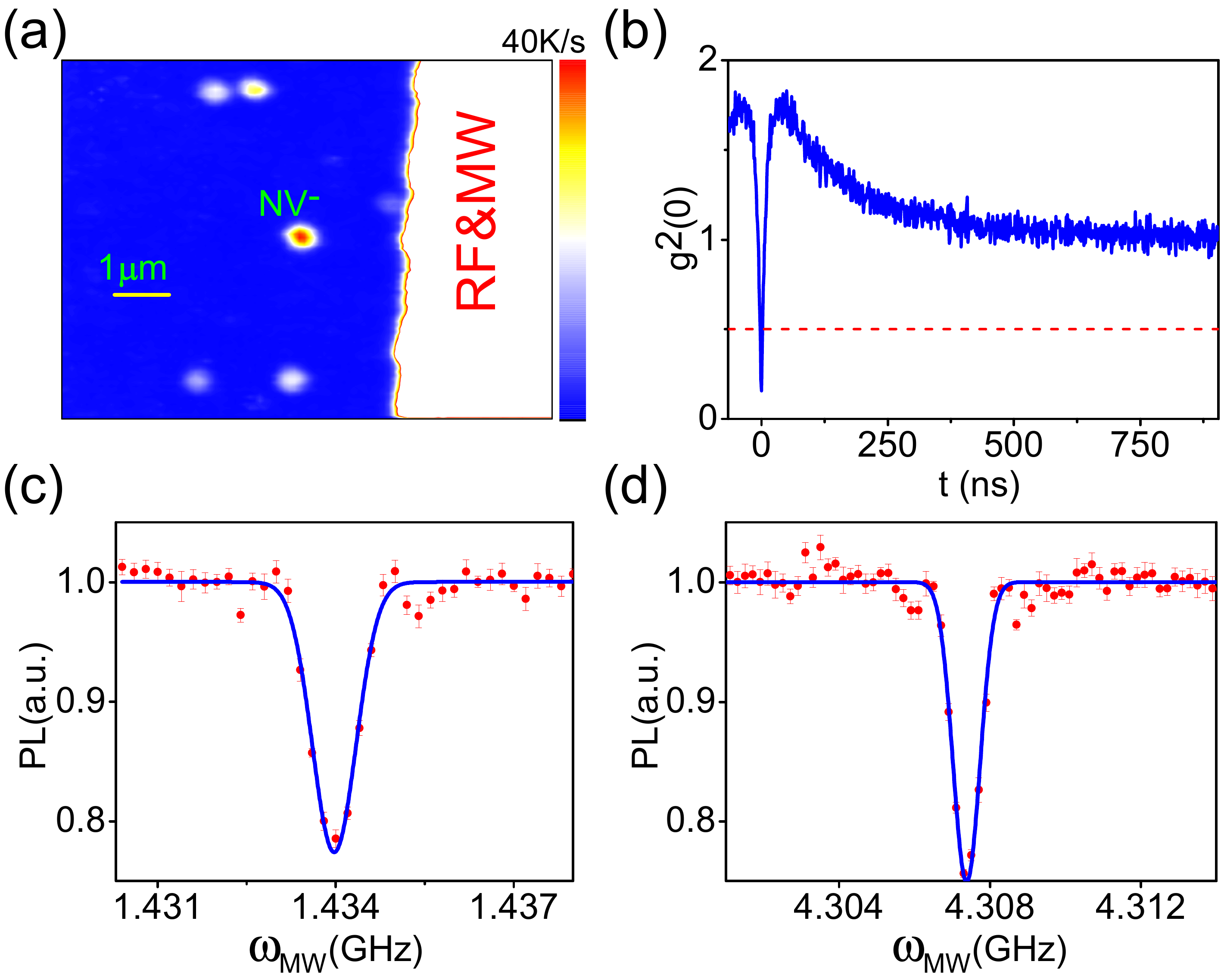}
\caption{(a) Confocal image of the NV center used in the experiment with a signal-to-noise ratio of 140:1. A coplanar waveguide antenna was deposited to deliver microwave pulses to the NV center. (b) Fluorescence correlation function. (c)-(d) ODMR spectra for the single NV center. The transition frequencies for $\left| 0 \right\rangle  \to \left| 1 \right\rangle $ and $\left| 0 \right\rangle  \to \left| 2 \right\rangle $ are ${\omega _{0, -1}}={\text{1}}{\text{.43398(1)}}$ $GHz $ and ${\omega _{0, 1}}={\text{4}}{\text{.30738(1)}}$ $GHz$, respectively.}
\label{SFig2S}
\end{figure}

Single photon emission from a single NV center was verified by measuring the photon correlation function ${{\text{g}}^2}{\text{(}}\tau )$ as shown in Fig.\ref{SFig2S}(b). And ${{\text{g}}^2}{\text{(}}\tau ) < 0.5$ indicates a single NV center. To form a simple V-type three-level system, a magnetic field of $51$ $mT$ was applied along the NV axis using a permanent magnet. Under this condition, the flip-flop process between electron-spin and nuclear-spin during optical pumping \cite{NVDNP} leads to polarizing the nitrogen nuclear spin of NV center after $3$ $\mu s$  green laser illumination. The Zeeman energy from the $51$ $ mT$ magnetic field shifts the respective energy differences between $\left| 0 \right\rangle \equiv \left| {{m_s} =  0} \right\rangle  \leftrightarrow \left| 1 \right\rangle \equiv \left| {{m_s} =  -1} \right\rangle$  and $\left| 0 \right\rangle  \leftrightarrow \left| { 2} \right\rangle \equiv \left| {{m_s} =  1} \right\rangle$ from the zero-field splitting, $2.870 $ $GHz$, to $1.43398(1)$ $ GHz$ and $4.30738(1) $ $GHz$, as shown in Fig.\ref{SFig2S}(c)-\ref{SFig2S}(d).

\subsection{Decoherence time}
\begin{figure}[tbp]
\centering
\includegraphics[width=8cm]{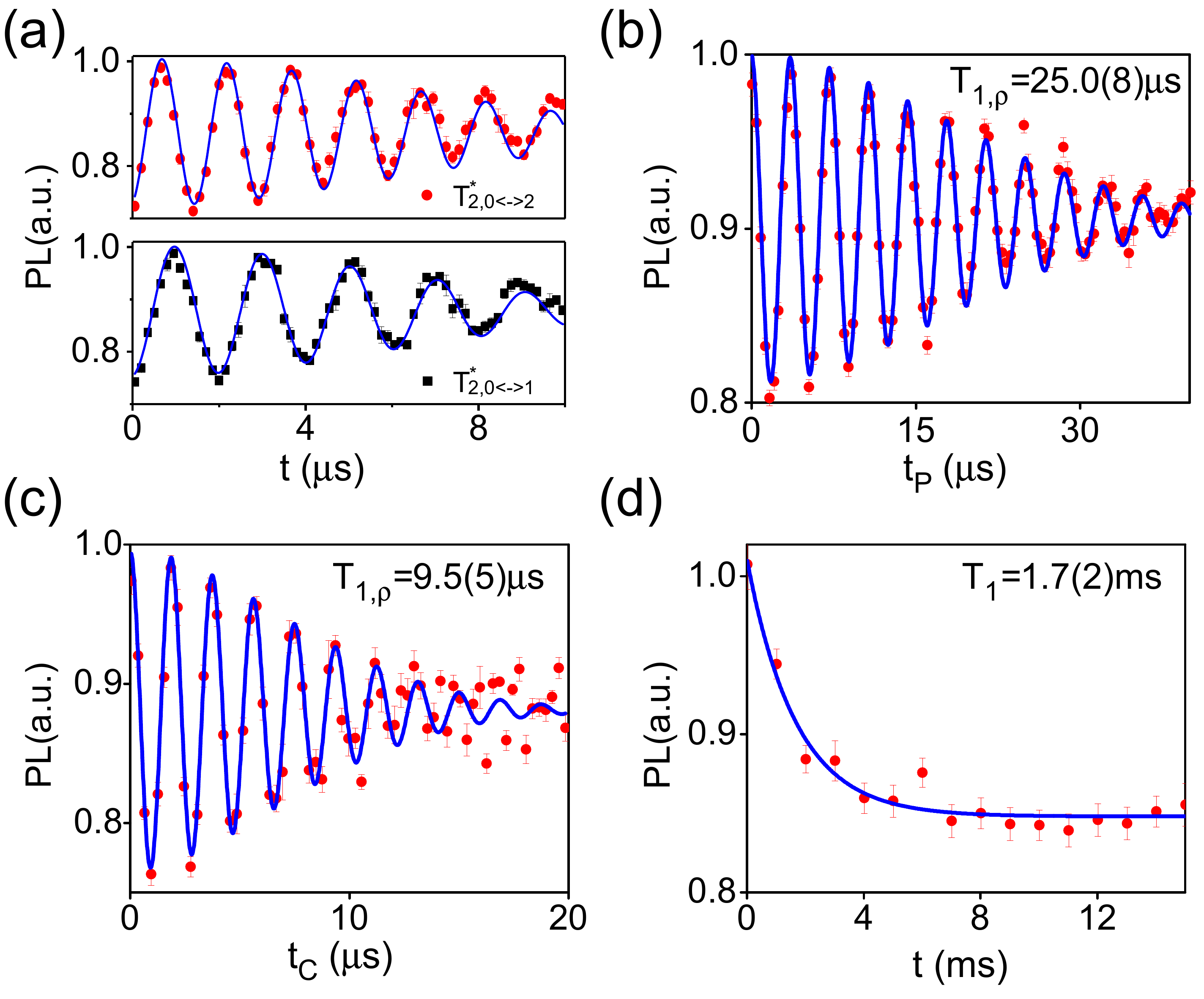}
\caption{(a) Result of the Ramsey experiment for the electron spin of NV center. Black squares and red circles correspond to $\left| 0 \right\rangle  \leftrightarrow \left| 1 \right\rangle $ and $\left| 0 \right\rangle  \leftrightarrow \left| 2 \right\rangle $, respectively. Experimental data are fitted by $y(t) = a\exp \left[ { - {{\left( {t/T_2^*} \right)}^2}} \right]\cos (2\pi \omega t)+b$ and denoted with solid blue curves. The dephasing time of NV center was measured to be $T_{2,0 \leftrightarrow 1}^* ={\text{8}}{\text{.2(3)}}$ $\mu s$ and $T_{2,0 \leftrightarrow 2}^* = {\text{8}}{\text{.7(4)}}$ $\mu s$. (b)-(c) Rabi oscillation of the electron spin between ground state sublevels of NV center. The experimental data (red dots) was fitted by a damped sine function (blue curves) written as $y(t)  =  a{\exp \left[{ - {{(t/{T_{1,\rho }})}^2}}\right ]}{\cos}\left( {\pi \frac{{x - {x_c}}}{w}} \right) + b$. We can get ${T_{1,\rho }} = 25.0(8)$ $\mu s$ ($9.5(5)$ $\mu s$) for only the probe (coupling) field. (d) The depolarization process of NV center
(red dots) was fitted by $y(t) = a{\exp \left({ - t/{T_1}}\right)} + b$ and we got ${T_1} = 1.7(2)$ $ms$.}
\label{Sfig3}
\end{figure}

For single NV center, the electron spin states dephasing \cite{QDP1,QIP1,NVH,NVT12} is the main part of quantum decoherence and has an impact on the ATS. The dephasing time of NV center was measured by the Ramsey interferometer \cite{NVH,NVT12}. By fitting experimental data as shown in Fig.S3(a), we got $T_{2,0 \leftrightarrow 1}^* = {\text{8}}{\text{.2(3)}}$ $\mu s$ and $T_{2,0 \leftrightarrow 2}^* ={\text{8}}{\text{.7(4)}}$ $\mu s$. We also measured the resonant Rabi oscillation of NV center driven only by the probe or coupling field as shown in Fig.\ref{Sfig3}(b)-(c). The Rabi frequency was kept same as in Fig.\ref{fig2} and ${\Omega _c}/{\Omega _p} = 2$. The damping times are ${T_{1,\rho }} = 25.0(8)$ $\mu s$ and ${T_{1,\rho }} = 9.5(5)$ $\mu s$ for $\left| 0 \right\rangle  \leftrightarrow \left| 1 \right\rangle $ and $\left| 0 \right\rangle  \leftrightarrow \left| { 2} \right\rangle $, respectively, by fitting the data. After the depolarization time of NV center was measured as shown in Fig.\ref{Sfig3}(d) with $T_1 =1.7(2)$ $ms$, we confirmed that the
dephasing process, which caused by nuclear spin bath, dominated the quantum decoherence of NV center.

\section{Theoretical model of the ATS with quantum state manipulation}

The Hamiltonian for NV center under driving fields in experiment reads
\begin{equation}
\begin{aligned}
H &= {\omega _{0,-1}}\left| 1 \right\rangle \langle 1| +{\omega _{0, 1}}\left| 2 \right\rangle \langle 2|
\\&+ {\Omega _c}\cos ({\omega _c}t + {\varphi _c})\left[ {\left| 1 \right\rangle \langle 0| + \left| 0 \right\rangle \langle 1|} \right]
\\&+ {\Omega _p}\cos ({\omega _p}t + {\varphi _p})\left[ {\left| 2 \right\rangle \langle 0| + \left| 0 \right\rangle \langle 2|} \right]\text{,}
\end{aligned}
\end{equation}
where $\varphi _c$ and $\varphi _p$ are the initial phases of the coupling and probe fields, respectively.

After transforming to a frame co-rotating with the two driving fields via ${U_0} = {e^{i{H_0}t}}$ with ${H_0} = {\omega _c}\left| 1 \right\rangle \left\langle 1 \right| + {\omega _p}\left| 2 \right\rangle \left\langle 2 \right|$, we get
\begin{equation}
\begin{aligned}
H = \left( {\begin{array}{*{20}{c}}
  0&{\frac{{{\Omega _c}}}{2}{e^{i{\varphi _c}}}}&{\frac{{{\Omega _p}}}{2}{e^{i{\varphi _p}}}} \\
  {\frac{{{\Omega _c}}}{2}{e^{ - i{\varphi _c}}}}&{{\Delta _c}}&0 \\
  {\frac{{{\Omega _p}}}{2}{e^{ - i{\varphi _p}}}}&0&{{\Delta _p}}
\end{array}} \right)\text{.}
\end{aligned}
\label{H1}
\end{equation}%

\subsection{The neglect of the initial phases of driving fields}
After initializing the NV center into its ground state $\left| 0 \right\rangle $, we can apply quantum gate $U = {e^{ - iHt}}$ operation.
Hence, $\left| \psi  \right\rangle  = {e^{ - iHt}}\left| 0 \right\rangle$. Finally, we detect the final state, which does not distinguish the states $\left| 1 \right\rangle  \equiv \left| {{m_s} = -1} \right\rangle$ and $ \left| 2 \right\rangle  \equiv \left| {{m_s} =  1} \right\rangle $. So we can use the measurement operator $\left| 0 \right\rangle \left\langle 0 \right|$ to describe the detection process

\begin{equation}
\begin{aligned}
{P_{\left| 0 \right\rangle \left\langle 0 \right|}} &= tr\left( {\left| 0 \right\rangle \left\langle 0 \right|\left| \psi  \right\rangle \left\langle \psi  \right|} \right)
\\& = \left\langle 0 \right|\left| \psi  \right\rangle \left\langle \psi  \right|\left| 0 \right\rangle
\\& = {\left| {\left\langle 0 \right|V{V^\dag }{e^{ - iHt}}V{V^\dag }\left| 0 \right\rangle } \right|^2}.
\end{aligned}
\label{P1}
\end{equation}%
Just letting $V = \left( {\begin{array}{*{20}{c}}
  1&0&0 \\
  0&{{e^{ - i{\varphi _c}}}}&0 \\
  0&0&{{e^{ - i{\varphi _p}}}}
\end{array}} \right)$, we have
\begin{equation}
\begin{aligned}
{V^\dag }HV &= \left( {\begin{array}{*{20}{c}}
  0&{\frac{{{\Omega _c}}}{2}}&{\frac{{{\Omega _p}}}{2}} \\
  {\frac{{{\Omega _c}}}{2}}&{{\Delta _c}}&0 \\
  {\frac{{{\Omega _p}}}{2}}&0&{{\Delta _p}}
\end{array}} \right)
\\& \triangleq \bar H.
\end{aligned}
\label{H2}
\end{equation}%
By substituting Eq.(\ref{H1}) and Eq.(\ref{H2}) to Eq.(\ref{P1}), we get
\begin{equation}
\begin{aligned}
 {P_{\left| 0 \right\rangle \left\langle 0 \right|}} = {\left| {\left\langle 0 \right|{e^{ - i\bar Ht}}\left| 0 \right\rangle } \right|^2},
\end{aligned}
\end{equation}%
which means that the arbitrary initial phases of driving fields do not have any effect and can be neglected in the experiment.\par

\subsection{The optimal demonstration of ATS}

\begin{figure}[tp]
\centering
\includegraphics[width=8cm]{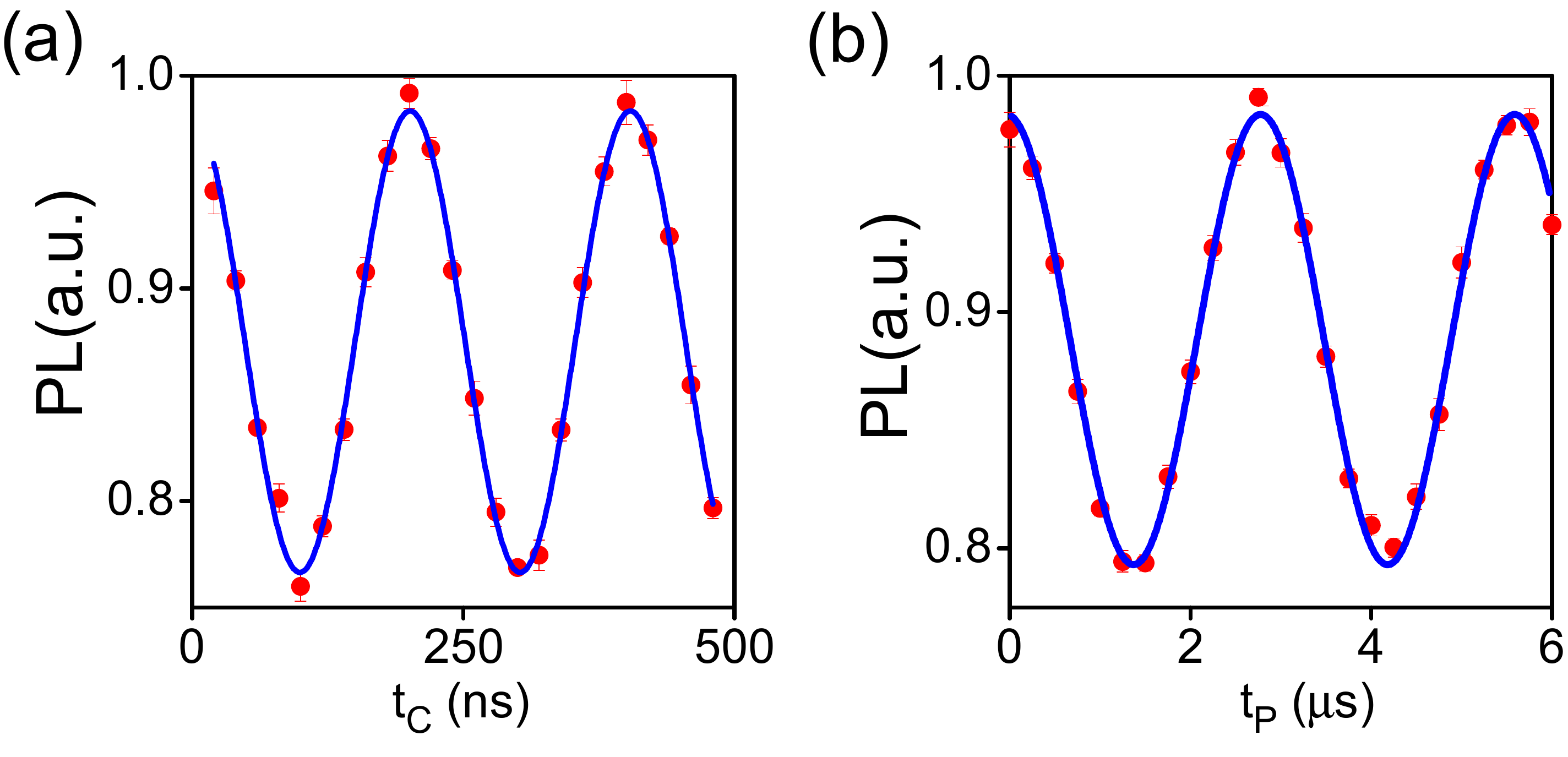}
\caption{(a)-(b) Rabi oscillation of the coupling and probe field for investigating dynamical process of ATS in main text Fig.3a-d and ${\Omega _c}/{\Omega _p} = 14$.}
\label{Sfig4}
\end{figure}

Now we chose the coupling field resonant with the single NV center. And in the dressed-state picture, Eq.(\ref{H1}) (omitting the initial phases) can be expressed as
\begin{equation}
\begin{aligned}
{H_d} = \left( {\begin{array}{*{20}{c}}
  {\frac{{{\Omega _c}}}{2}}&0&{\frac{{\sqrt 2 {\Omega _p}}}{4}} \\
  0&{ - \frac{{{\Omega _c}}}{2}}&{\frac{{\sqrt 2 {\Omega _p}}}{4}} \\
  {\frac{{\sqrt 2 {\Omega _p}}}{4}}&{\frac{{\sqrt 2 {\Omega _p}}}{4}}&{{\Delta _p}},
\end{array}} \right) \text{,}
\end{aligned}
\label{Hd}
\end{equation}%
where $\left|  +  \right\rangle  = \frac{{\left| 0 \right\rangle  + \left| 1 \right\rangle }}{{\sqrt 2 }}$,
$\left|  -  \right\rangle  = \frac{{\left| 0 \right\rangle  - \left| 1 \right\rangle }}{{\sqrt 2 }}$
and $\left| 2 \right\rangle  = \left| 2 \right\rangle $ form new basis vectors. The frequency of the probe field was scanned to make sure that
the Rabi frequency of probe field is $1/14$ of that of the coupling field, as shown in Fig.\ref{Sfig4}(a)-(b).\par

Case I:  ${\Delta _p} = 0$, due to the large detuning between energy level of the NV center, the effect of the probe field can be neglected. The spin state is in bright state $\left| 0 \right\rangle $ when ${\Omega _c}t = 2n\pi ,n = 1,2,3 \cdots $, as shown in Fig.\ref{fig2}(b) in the main text.

Case II:  ${\Delta _p}  \approx \frac{{{\Omega _c}}}{2}$, the middle energy level effect can be eliminated by the second order perturbation theory with
\begin{equation}
\begin{aligned}
{H_d} = \left( {\begin{array}{*{20}{c}}
  {\frac{{{\Omega _c}}}{2}}&0&{\frac{{\sqrt 2 {\Omega _p}}}{4}} \\
  0&{ - \frac{{{\Omega _c}}}{2}}&0 \\
  {\frac{{\sqrt 2 {\Omega _p}}}{4}}&0&{{\Delta _p} + \frac{{\Omega _p^2}}{{8{\Omega _c}}}}
\end{array}} \right).
\end{aligned}
\label{Hd2}
\end{equation}%

So when the frequency of the probe field was scanned and once ${\Delta _p} = \frac{{{\Omega _c}}}{2} - \frac{{\Omega _p^2}}{{8{\Omega _c}}}$, we have
\begin{equation}
\begin{aligned}
{P_{\left| 0 \right\rangle \langle 0|}} &= {\left| {\frac{1}{{\sqrt 2 }}\left( {\begin{array}{*{20}{c}}
  1&1&0
\end{array}} \right){e^{ - i{H_d}t}}\frac{1}{{\sqrt 2 }}\left( {\begin{array}{*{20}{c}}
  1 \\
  1 \\
  0
\end{array}} \right)} \right|^2}
\\&= {\left| {\frac{1}{{\sqrt 2 }}\left( {\begin{array}{*{20}{c}}
  1&1&0
\end{array}} \right)\frac{1}{{\sqrt 2 }}\left( {\begin{array}{*{20}{c}}
  {\cos \frac{{\sqrt 2 {\Omega _p}t}}{4}} \\
  {{e^{i{\Omega _c}t}}} \\
  { - i\sin \frac{{\sqrt 2 {\Omega _p}t}}{4}}
\end{array}} \right)} \right|^2}
\\&= \frac{1}{4}{\left| {\cos \frac{{\sqrt 2 {\Omega _p}t}}{4} + {e^{i{\Omega _c}t}}} \right|^2}.
\end{aligned}
\end{equation}
To get the highest contrast for the ATS, the conditions are:
\begin{equation}
{\Omega _{p}}t =2\sqrt{2}(2k-1)\mathrm{\pi }\text{, }
{\Omega _{c}}t =2{n}\mathrm{\pi }\text{,}
\end{equation}
or
\begin{equation}
{\Omega _{p}}t =2\sqrt{2}(2k)\mathrm{\pi }\text{, }
{\Omega _{c}}t =(2n-1)\mathrm{\pi }\text{,}
\end{equation}
where $n, k= 1,2,3 \cdots$.

Case III: ${\Delta _p}  \approx -\frac{{{\Omega _c}}}{2}$, the situation is similar to Case II. So the conditions for the observation of the highest contrast is same and the resonance frequency of the probe field is ${\Delta _p} =  - \frac{{{\Omega _c}}}{2} + \frac{{\Omega _p^2}}{{8{\Omega _c}}}$.\par

\subsection{The ATS with non-resonant driving fields}
For  ${\Omega _{c}\gg \Omega _{p}}$, the ATS can be expressed as
\begin{equation}
\begin{aligned}
{\Delta _{AT}} \approx {\Omega _c} - \frac{{\Omega _p^2}}{{4{\Omega _c}}} \approx {\Omega _c}.
\end{aligned}
\end{equation}%
If the coupling field is not resonant, we have
\begin{equation}
\begin{aligned}
\bar H &= \frac{{{\Delta _c} + {\Omega _{eff}}}}{2}\left|  +  \right\rangle \left\langle  +  \right| + \frac{{{\Omega _c}{\Omega _p}}}{{2\sqrt {2\Omega _{eff}^2 + 2{\Delta _c}{\Omega _{eff}}} }}\left|  +  \right\rangle \left\langle 2 \right|
\\&+ \frac{{{\Omega _c}{\Omega _p}}}{{2\sqrt {2\Omega _{eff}^2 + 2{\Delta _c}{\Omega _{eff}}} }}\left| 2 \right\rangle \left\langle  +  \right| + \frac{{{\Delta _c} - {\Omega _{eff}}}}{2}\left|  -  \right\rangle \left\langle  -  \right|
\\&+ \frac{{{\Omega _c}{\Omega _p}}}{{2\sqrt {2\Omega _{eff}^2 - 2{\Delta _c}{\Omega _{eff}}} }}\left|  -  \right\rangle \left\langle 2 \right|
\\&+ \frac{{{\Omega _c}{\Omega _p}}}{{2\sqrt {2\Omega _{eff}^2 - 2{\Delta _c}{\Omega _{eff}}} }}\left| 2 \right\rangle \left\langle  -  \right| + \frac{{{\Delta _p}}}{2}\left| 2 \right\rangle \left\langle 2 \right|
\end{aligned}
\end{equation}%
in the bases of $\left|  +  \right\rangle  = \frac{{{\Omega _c}}}{{\sqrt {2\Omega _c^2 + 2\Delta _c^2 + 2{\Delta _c}\sqrt {\Omega _c^2 + \Delta _c^2} } }}\left| 0 \right\rangle + \frac{{{\Delta _c} + \sqrt {\Omega _c^2 + \Delta _c^2} }}{{\sqrt {2\Omega _c^2 + 2\Delta _c^2 + 2{\Delta _c}\sqrt {\Omega _c^2 + \Delta _c^2} } }}\left| 1 \right\rangle $ and $\left|  -  \right\rangle  = \frac{{{\Omega _c}}}{{\sqrt {2\Omega _c^2 + 2\Delta _c^2 - 2{\Delta _c}\sqrt {\Omega _c^2 + \Delta _c^2} } }}\left| 0 \right\rangle  + \frac{{{\Delta _c} - \sqrt {\Omega _c^2 + \Delta _c^2} }}{{\sqrt {2\Omega _c^2 + 2\Delta _c^2 - 2{\Delta _c}\sqrt {\Omega _c^2 + \Delta _c^2} } }}\left| 1 \right\rangle $. Hence, when the two resonant transitions have non-zero detuning, the splitting exhibits asymmetric ATS (unequal transmission dips) and ${\Delta _{AT}} \approx {\Omega _{eff}}=\sqrt {\Omega _c^2 + \Delta _c^2}.$

\subsection{The failure in the observation of ATS in a dephasing dominated system with traditional method}

The traditional method to observe the ATS is based on the distribution of the population of static state under long-pulse driving fields \cite{ATS1,ATS2,ATQD1,ATM1,AAT1,AAT2,AAT3,AAT4,AAT5,AAT6,AAT7}, which can not be applied to demonstrate the ATS in the dephasing dominated system. Here, we employ Lindblad equation for the steady-state solution with
\begin{equation}
\begin{aligned}
{\dot \rho ^I} =  - i[{H_I},{\rho ^I}] + \sum\limits_j {D(A_j^I){\rho ^I}} ,
\end{aligned}
\label{LE}
\end{equation}%
where $D({A^I}){\rho ^I} = A{\rho ^I}{A^\dag } - \left\{ {{A^\dag }A,{\rho ^I}} \right\}/2$ and ${H_I} = \bar H$. The longitudinal relaxation from $i$ to $j$ can be written as ${A_{diss}} = \sqrt {{\Gamma _{ij}}} \left| j \right\rangle \left\langle i \right|$
and the dephasing process for state $i$ is ${A_{de}} = \sqrt {2{\gamma _a}} \left| a \right\rangle \left\langle a \right|$.\par
For the dephasing channels,
\begin{equation}
\begin{aligned}
D({A_{de}})\rho  = \left[ {\begin{array}{*{20}{c}}
  0&{ - {\gamma _1}{\rho _{01}}}&{ - {\gamma _2}{\rho _{02}}} \\
  { - {\gamma _1}{\rho _{10}}}&0&{ - {\gamma _3}{\rho _{12}}} \\
  { - {\gamma _2}{\rho _{20}}}&{ - {\gamma _3}{\rho _{21}}}&0
\end{array}} \right].
\end{aligned}
\end{equation}%
Just letting ${\rho _{ij}} \to \rho _{ij}^I$ for the expression of the dephasing process, we can transform the lab frame to the rotation frame.\par
In the NV center system, the dephasing process is much faster than the longitudinal relaxation($\gamma \gg  \Gamma $). So we can omit the longitudinal relaxation and get equation for the steady state after long-pulse driving,
\begin{equation}
\begin{aligned}
0 =  - i[{H_I},{\rho _I}] + D({A_{de}}){\rho_I}.
\end{aligned}
\end{equation}
Hence, $\rho _{12}^I = \rho _{01}^I = \rho _{02}^I = 0$.
At last,
\begin{equation}
\begin{aligned}
\rho _{00}^I = \rho _{11}^I = \rho _{22}^I = \frac{1}{3},
\end{aligned}
\end{equation}
which means that the ATS or EIT cannot be observed with the traditional observation method. And it also holds for cascade and $\Lambda $ three level system to investigate EIT or ATS based on the traditional observation method in the dephasing dominated quantum decoherence system.\par
\begin{figure}[bp]
\centering
\includegraphics[width=8cm]{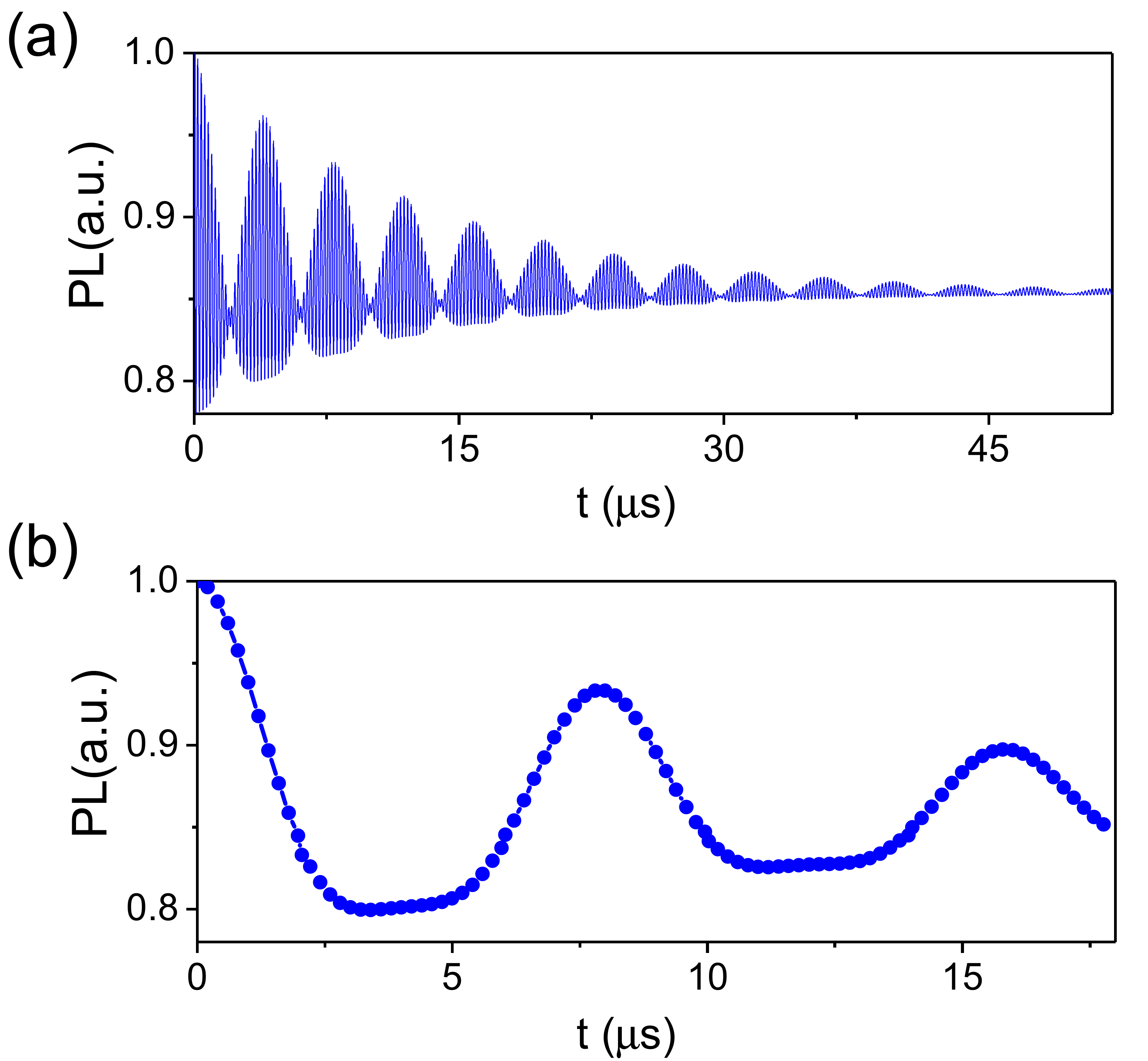}
\caption{(a)-(b) The result of the dynamical process of
ATS shown with blue line based on solving the Lindblad equation. The parameters used in this figure are ${\Omega _c} = 5$ MHz, ${\Omega _c}/{\Omega _p} = 14$, ${\gamma _1} = {\gamma _2} = 0.0784$ MHz, ${\gamma _3} = 2{\gamma _2}$, ${\Delta _c} = 0, {\Delta _p} = \frac{{{\Omega _c}}}{2} - \frac{{\Omega _p^2}}{{8{\Omega _c}}}$.}
\label{Sfig5}
\end{figure}
\subsection{The dynamical process of ATS and simulation}
The dynamical process of ATS in NV center can also be numerically simulated with the Lindblad equation. The independent equations of Eq.(\ref{LE}) are
\begin{equation}
\begin{gathered}
  \dot \rho _{00}^I =  - i\left[ {\frac{{{\Omega _c}}}{2}(\rho _{10}^I - \rho _{01}^I) + \frac{{{\Omega _p}}}{2}(\rho _{20}^I - \rho _{02}^I)} \right] \text{,}\hfill \\
  \dot \rho _{01}^I =  - i\left[ {\frac{{{\Omega _c}}}{2}(\rho _{11}^I - \rho _{00}^I) + \frac{{{\Omega _p}}}{2}\rho _{21}^I} \right] - {\gamma _1}\rho _{01}^I \text{,}\hfill \\
  \dot \rho _{02}^I =  - i\left[ {\frac{{{\Omega _c}}}{2}\rho _{12}^I + \frac{{{\Omega _p}}}{2}(\rho _{22}^I - \rho _{00}^I) - {\Delta _p}\rho _{02}^I} \right] - {\gamma _2}\rho _{02}^I \text{,}\hfill \\
  \dot \rho _{11}^I =  - i\left[ {\frac{{{\Omega _c}}}{2}(\rho _{01}^I - \rho _{10}^I)} \right] \text{,}\hfill \\
  \dot \rho _{12}^I =  - i\left[ {\frac{{{\Omega _c}}}{2}\rho _{02}^I - \frac{{{\Omega _p}}}{2}\rho _{10}^I - {\Delta _p}\rho _{12}^I} \right] - {\gamma _3}\rho _{12}^I \text{,}\hfill\\
  1 = \rho _{00}^I + \rho _{11}^I + \rho _{22}^I \text{.}\hfill \\
\end{gathered}
\end{equation}
The last equation is the additional constraint of completeness. Just letting
\begin{equation}
\begin{gathered}
  \rho _{00}^I = {y_1} \text{,}\hfill\\
  \rho _{11}^I = {y_2} \text{,}\hfill\\
  \rho _{01}^I = {y_3} + i{y_4} \hfill \text{,}\\
  \rho _{02}^I = {y_5} + i{y_6} \hfill \text{,}\\
  \rho _{12}^I = {y_7} + i{y_8}\hfill \text{,} \\
\end{gathered}
\end{equation}
we can convert the physical equation to the linear ordinary differential equations and solve them with the Runge-Kutta method. The fluorescence intensity of the NV center is given by $I = 1 - C + C{y_1}$ with $C=PL_{{m_{s}}=0}-PL_{{m_{s}}=\pm 1}=0.22$ is fluorescence contrast for different spin states \cite{NVDNP}. Fig.\ref{Sfig5}(a) shows the result of the time dependence of the probabilities for a particular set of conditions beginning with $\left| 0 \right\rangle $. When the duration time of both driving fields is larger than the dephasing time, the system will become a maximally mixed state.

If the duration time of the driving fields satisfies
$\Omega _{c}t =2{n}\mathrm{\pi }$ ($n=1,2,3,\cdots $), the envelope line will be obtained as shown in Fig.\ref{Sfig5}(b). There is a little discrepancy between the theory and experimental result as shown in Fig.\ref{fig3}(c) in the main text. The most important factors causing the deviation would be the environment treatment of NV center. For the present sample, the decoherence of NV center is dominated by the hyperfine interaction with the $^{13}C$ nuclear spins, which form a nuclear spin bath. The bath spins involved in the decoherence of NV center is much more complicated than those in quantum dots and shallow donors \cite{NVDE}.

\end{document}